\pdfoutput=1
\documentclass[conf]{new-aiaa}
\usepackage[margin=1in]{geometry}               
\usepackage{graphicx}
\usepackage[utf8]{inputenc}
\usepackage[english]{babel}
\usepackage{amsfonts, amsmath, latexsym, verbatim, xspace, setspace}
\usepackage{comment}
\usepackage{listings}
\usepackage{caption}
\usepackage{subcaption} 
\usepackage[ruled,vlined]{algorithm2e}
\usepackage{harpoon}
\usepackage{color}
\usepackage{soul}

\definecolor{mygreen}{RGB}{28,172,0} 
\definecolor{mylilas}{RGB}{170,55,241}

\usepackage{epstopdf}
\usepackage{calc}
\usepackage[outside]{coordsys}
\usepackage{tikz}
\usepackage{pgfplots}
\usetikzlibrary{plotmarks,shapes, calc, arrows, backgrounds, decorations.markings}
\tikzstyle{vertex}=[circle, draw, inner sep=0pt, minimum size=6pt]
\usepgfplotslibrary{polar}
\usepgflibrary{shapes.geometric}
\tikzstyle{place}=[circle,draw=blue!50,fill=blue!20,thick,
                   inner sep=0pt,minimum size=6mm]
\tikzstyle{transition}=[circle,draw=black!50,fill=black!20,thick,
                        inner sep=0pt,minimum size=4mm]
                        
\usepackage{multirow}
\usepackage{tabularx} 
\newcolumntype{Y}{>{\centering\arraybackslash}X}
\newcolumntype{P}[1]{>{\centering\arraybackslash}p{#1}}
\usepackage{arydshln} 
\usepackage{threeparttable}
\usepackage{makecell}
\pgfplotsset{compat=1.16}

\usepackage[capitalise]{cleveref}
\Crefformat{equation}{Eq.~(#2#1#3)}
\Crefformat{equation}{Equation.~(#2#1#3)}
\Crefmultiformat{equation}{Eqs.~(#2#1#3)}{ and~(#2#1#3)}{, (#2#1#3)}{ and~(#2#1#3)}
\Crefmultiformat{equation}{Equations~(#2#1#3)}{ and~(#2#1#3)}{, (#2#1#3)}{ and~(#2#1#3)}
\Crefrangeformat{equation}{Eqs.~(#3#1#4) to~(#5#2#6)}
\Crefrangeformat{equation}{Equations~(#3#1#4) to~(#5#2#6)}
\Crefformat{figure}{Fig.~#2#1#3}
\Crefformat{figure}{Figure~#2#1#3}
\Crefmultiformat{figure}{Figs.~#2#1#3}{ and~#2#1#3}{, #2#1#3}{ and~#2#1#3}
\Crefmultiformat{figure}{Figures~#2#1#3}{and~#2#1#3}{, #2#1#3}{ and~#2#1#3}
\Crefrangeformat{figure}{Fig.~#3#1#4 to~#5#2#6}
\Crefrangeformat{figure}{Figures~#3#1#4 to~#5#2#6}
\addtocounter{figure}{0}

\newtheorem{theorem}{Theorem}

\numberwithin{equation}{section}

\title{An Unstructured Mesh Approach to Nonlinear Noise Reduction for Coupled Systems\footnote{Distribution A: Approved for Public Release; Distribution is Unlimited.
PA No. AFRL-2022-4193}}
\author{Aaron Kirtland}\affil{Cavendish Laboratories, Cavendish, VT, USA}
\author{Jonah Botvinick-Greenhouse\footnote{Corresponding author: \texttt{jrb482@cornell.edu}}}\affil{Center for Applied Mathematics, Cornell University, Ithaca, NY 14850, USA}
\author{Marianne DeBrito}\affil{University of Michigan, Ann Arbor, MI 48109, USA}
\author{Megan Osborne}\affil{Rensselaer Polytechnic Institute, Troy, NY 12180, USA}
\author{Casey Johnson}\affil{University of California, Los Angeles, CA, USA}
\author{Robert S. Martin}\affil{DEVCOM-ARL U.S. Army Research Office, USA}
\author{Samuel J. Araki}\affil{Jacobs Technology Inc., Edwards Air Force Base, CA 93524, USA}
\author{Daniel Q. Eckhardt}\affil{In-Space Propulsion Branch, Air Force Research Laboratory, Edwards Air Force Base, CA 93524, USA}

\begin{document}
\maketitle
\begin{abstract}
	To address noise inherent in electronic data acquisition systems and real world sources, Araki et al. [Physica D: Nonlinear Phenomena, 417 (2021) 132819] demonstrated a grid based nonlinear technique to remove noise from a chaotic signal, leveraging a clean high-fidelity signal from the same dynamical system and ensemble averaging in multidimensional phase space. This method achieved denoising of a time-series data with 100\% added noise but suffered in regions of low data density. To improve this grid-based method, here an unstructured mesh based on triangulations and Voronoi diagrams is used to accomplish the same task. The unstructured mesh more uniformly distributes data samples over mesh cells to improve the accuracy of the reconstructed signal. By empirically balancing bias and variance errors in selecting the number of unstructured cells as a function of the number of available samples, the method achieves asymptotic statistical convergence with known test data and reduces synthetic noise on experimental signals from Hall Effect Thrusters (HETs) with greater success than the original grid-based strategy. 
\end{abstract}

\section{Introduction}
Real-world data obtained from complex physical systems suffers from various types of uncertainties that are broadly categorized as either epistemic or aleatory. The latter occurs as the result of natural random processes, while the former occurs as a result of uncertainty introduced by the model through which the data is viewed. While noise in measurements of physical systems is generally thought of as aleatory, the source of this type of uncertainty can result from high frequency system dynamics unresolvable by the measurement device or as introduced by the measurement device itself. The distinction between noise and unresolved dynamics is therefore a subtle question further complicated by the boundary drawn between the system being studied, the external environment, and the instrument used to perform the study.  For rocket devices in general, due to the immense complexity of the physical processes (e.g. chemical combustion or plasma dynamics) and timescales involved, it is generally difficult to interpret what signal components constitute measurement noise versus what corresponds to high frequency dynamics with unknown causes. Even with over 100 years of rocket propulsion, this physical complexity makes both the interpretation of the data collected from these devices and the detailed prediction of their behavior extremely challenging. 

This work uses an electrostatic plasma propulsion device, the Hall-effect thruster (HET), to exemplify the challenges posed by the afore mentioned uncertainities. HETs are plasma devices that electrostatically accelerate ionized gas to produce thrust, and as such have tightly coupled particles--neutrals, ions and electrons--all evolving at vastly different timescales to efficiently produce thrust. Because electron dynamics occur at much faster time scales than ions and neutrals, their dynamic fluctuations can easily be mistaken for noise, yet their collective behavior dramatically influences the performance of the thruster. More to this, since these devices operate in  high vacuum ($<10^{-5}$ torr), measurements occur across long wires which introduce impedance and noise, which can be difficult to distinguish from the plasma dynamics. This is not to discount the various techniques and models that exist to study devices such as these \cite{sauer1992noise,grassberger1993noise,hammel1990noise,farmer1991optimal,sternickel2001nonlinear,moore2015improvements}, but rather to address some of the difficult plasma physics that eludes the community to this day. New approaches must be developed to better capture the true operation of these devices and realize their full capabilities through increased efficiency and predictability. 

The purpose of this paper is to extend the work done in \cite{araki2021grid}, to address mesh convergence issues observed due to bias errors, and to demonstrate the utility of the proposed technique when uncertainty about the noise is either aleatory or epistemic in the device data. The aim is to develop a denoising algorithm more capable of adapting to data density, which has the benefit of computational speedup and shows greater promise in distinguishing between noise and high-speed dynamics inherent to the system of interest. The approach taken utilizes an unstructured mesh constructed from the Voronoi diagram and incorporates linear interpolation between cell averages via the Delaunay triangulation. The paper is organized as follows. \Cref{sec:background} provides background and theory which is fundamental to this research. \Cref{sec:methods} details the proposed denoising algorithm. 
Sections \ref{sec:resultsLorenz} and \ref{sec:resultsHET} discuss results obtained for the Lorenz and HET systems.
Finally, \Cref{sec:conclusions} concludes the paper.

\section{Background and Theory}
\label{sec:background}
\subsection{Hall-Effect Thruster}
HETs are some of the most widely used solar electric propulsion devices on spacecrafts today, providing high efficiency propulsion for orbit raising and stationkeeping. HETs operate by electrostatically accelerating ionized plasma to high exhaust speed in order to produce thrust \cite{goebel2008fundamentals,boeuf2017tutorial}. 
A HET consists of an annular channel with an interior anode and a cathode placed externally to the channel. Inside the channel, the magnetic field primarily points radially outward to increase the electron confinement time, promoting ionization of the propellant gas. A HET can operate in either the quiescent mode or the breathing mode, but understanding the breathing mode is particularly important for improving the stability and the design of the thruster. This breathing operating mode oscillates almost periodically with slightly varying frequencies (i.e. quasi-periodic) and exhibits behavior resembling a limit cycle.

\subsection{Data Fusion Methods for HET}
A variety of time-resolved measurements have been taken to study the HET's oscillatory behavior, which often includes the high-fidelity discharge current with very high signal-to-noise ratio (SNR) as well as low-fidelity measurements inside the plasma. To better understand the dynamical system, previous researchers managed to improve the quality of the low-fidelity measurements by fusing multiple data sources and applying various techniques such as (1) the linear Fast Fourier Transform (FFT) decomposition~\cite{lobbia2010,lobbia2010thesis,durot2014,durot2016}, (2) a hardware-based filtering technique~\cite{biloiu2006,mazouffre2009,mazouffre2010,vaudolon2013,macdonald2012,macdonald2014}, and (3) a nonlinear reconstruction technique referred to as shadow manifold interpolation (SMI) ~\cite{eckhardt2019spatiotemporal, doi:10.1137/20M1350923}. 
While all of these methods yielded great success in studying the dynamics of a HET system, the FFT-based and hardware-based methods did not work for chaotic dynamics with non-smooth features. Though the SMI method did work in this situation, it required significant data storage and computational time.

To address the limitations found in the data fusion methods described above, Araki et al.~\cite{araki2021grid} demonstrated a new strategy for nonlinear denoising, leveraging the availability of high-fidelity data from one part of the system and effectively applying ensemble averaging in phase space for data in other parts of the system. The reconstruction method involved placing a uniform mesh over a time-delayed embedding of the clean high-fidelity reference signal as inspired by Takens' theorem~\cite{takens1981} and Convergent Cross-Mapping (CCM)~\cite{sugihara2012}. Although this method was shown to work in a chaotic system and the calculation was significantly faster than the SMI method, it showed poor reconstruction in regions of sparse data in phase space (e.g. near edge of the manifold). This was mitigated by applying a smoothing technique, but the approach was still insufficient in some cases. The reconstruction improved with an increased number of data points, but the convergence was slow as the error was primarily attributed to low data density regions. 

\subsection{Takens' Embedding Theorem}
In \cite{araki2021grid}, clean signals sampled from chaotic dynamical systems are used to recover information about the entire system's dynamics and reconstruct other signals which have been corrupted with Gaussian noise. These reconstructions are in part accomplished by the procedure of \textit{time-delay embedding}, in which the time-delays of a signal are used to construct a high-dimensional manifold capable of representing the full state of the system, up to a diffeomorphism. If the diffeomorphism can be accurately approximated, then the time-delays of a single state variable can be used to gain valuable insight into the dynamics of other state variables. This is precisely the relationship that facilitated the denoising of coupled signals in \cite{araki2021grid}. Takens' embedding theorem~\cite{takens1981} provides the theoretical justification for the time-delay embedding procedure and supplies conditions under which such a diffeomorphism is expected to exist. 


\begin{theorem}[Takens' Theorem \cite{takens1981}]
Let $\mathcal{M}$ be a compact manifold of dimension $m$.
For pairs $(\phi,y)$, where $\phi:\mathcal{M}\to\mathcal{M}$ is a smooth diffeomorphism and $y:\mathcal{M}\to \mathbb{R}$ is a smooth function, it is a generic property that $\Phi_{(\phi,y)}:\mathcal{M}\to\mathbb{R}^{2m+1}$, given by $\Phi_{(\phi,y)}(x)=(y(x),y(\phi(x)),…,y(\phi^{2m}(x)))$ is an embedding. Here, "smooth" means at least $C^2$.
\end{theorem}


\begin{figure}[!b]
    \centering
    \includegraphics[trim={0cm 6.5cm 0cm 6.5cm},clip,width = \textwidth]{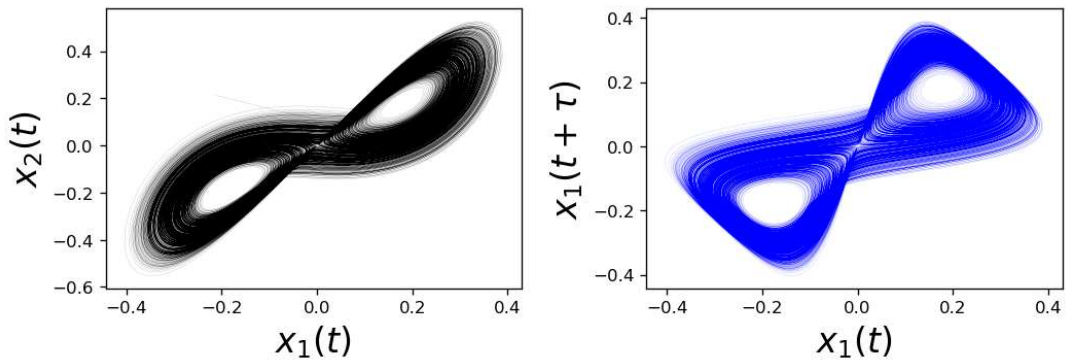}
    \caption{Visualization of the manifold $\mathcal{M}$ (left) and a shadow $\mathcal{M}^{x_1}$ (right) for the Lorenz system \eqref{eqn:Lorenz}. The correspondence between projections of the true dynamics and the time-delay embedded dynamics is clear.}
    \label{fig:lor_shadow}
\end{figure}

Under suitable assumptions, the diffeomorphism $\phi$ can be taken to be the time-$\tau$ flow map of a smooth vector field $v:\mathcal{M}\to \mathcal{M}$, for a time-delay $\tau > 0$. Therefore, when $X(t)$ is an observable of an autonomous dynamical process, Takens' theorem implies that the manifold $\mathcal{M}^X$ to which the time-lagged observations $(X(t),X(t+\tau),\dots,X(t+2m\tau))$ belong will be diffeomorphic to the true manifold $\mathcal{M}.$  Hereafter, $\mathcal{M}^X$ will be referred to as a \textit{shadow manifold}.
If two shadow manifolds, $\mathcal{M}^X$ and $\mathcal{M}^Y$, are created for observables $X$ and $Y$, respectively, both would share a diffeomorphic relationship with $\mathcal{M}$. Consequently, both $\mathcal{M}^X$ and $\mathcal{M}^Y$ would be related to one another via a diffeomorphism.

Because of this relationship, it is possible to use the time-indices of a collection of nearby points on $\mathcal{M}^X$ to locate a corresponding group of nearby points on $\mathcal{M}^Y$. In general, such points remain nearest on $\mathcal{M}^Y$ provided that the variables $X$ and $Y$ are causally related. Using this correspondence, knowledge of the variable $X$ can be used to forecast future states of dynamical trajectories on the manifold $\mathcal{M}^Y$. As the available training data increases, these predictions are expected to become more accurate. An ability to forecast the state on $\mathcal{M}^Y$ from points on $\mathcal{M}^X$ serves as evidence for a causal relationship between the variables $X$ and $Y$, which is the core idea of CCM \cite{sugihara2012}.
However, when one of the signals is corrupted by noise, the nearest neighbors on one shadow manifold instead correspond to a noisy ball of points on the other. By averaging over these noisy samples, the underlying cross map can be recovered. Thus, a clean signal can be used to achieve reconstructions of a causally related signal which has been corrupted with noise. These ideas provide the basis for the nonlinear noise reduction which was performed in \cite{araki2021grid}.




The similarities between the original dynamics and the time-delayed embedding can even be observed visually, as shown in \Cref{fig:lor_shadow}. Takens' theorem guarantees a minimal dimension for which the time-delayed embedding adequately represents the original manfiold. However, lower dimensional embeddings are often possible in practice.  Note that the selection of time lag and embedding dimension can impact the quality of the reconstructed signals, and that methods for optimally selecting these values are briefly described in \Cref{subsec:datapreparation}.



\subsection{Test Data}
\label{subsec:testdata}
In order to develop a denoising algorithm that accurately recovers the desired clean signal, the Lorenz system is first studied. 
The Lorenz system was chosen because it possesses similar dynamical characteristics to the time-delay embedded HET signals.
The differential equations governing the Lorenz system are given by
\begin{equation}
  \begin{array} {l} 
  \frac{dx_1}{dt} = \eta(x_2-x_1) \vspace{0.5ex}\\
  \frac{dx_2}{dt} = x_1(\alpha - x_3) - x_2  \vspace{0.5ex}\\
  \frac{dx_3}{dt} = x_1 x_2 - \beta x_3. 
  \end{array}
  \label{eqn:Lorenz}
\end{equation}
The values $\eta = 10, \alpha = 30,$ and $\beta = 8/3$ are used throughout the paper with initial conditions $x_1(0)=x_2(0)=x_3(0)=1$.

After demonstrating success of the proposed denoising technique on the Lorenz system, the algorithm is then applied to measurement data obtained from a sub-kW HET operating in a vacuum chamber. 
The thruster anode and cathode are independently connected to Pearson coils, and currents are measured in Amperes at a sampling frequency of 25~MHz.
Currents are also measured at different segments of metal rings of a cage that encloses the thruster.
Further details on the experimental setup and signals used are provided in \cite{macdonald2016,eckhardt2019spatiotemporal}.
It is assumed that the extrinsic noise has been mostly removed in pre-processing the HET signals, and as such they are considered to be "clean."

Throughout the experiments in \Cref{sec:resultsLorenz} and \Cref{sec:resultsHET}, synthetic Gaussian noise is added to a target signal, and another clean signal from the same system is used to reduce the noise level. Hereafter, let $\Tilde{Y}(t)$ denote the signal $Y(t)$ with additive Gaussian noise, such that $\Tilde{Y}(t)=Y(t)+\varepsilon(t)$. Specifically, at each sampling time $t_i$, it holds that $\varepsilon(t_i)\sim \mathcal{N}(0,\mu)$, where $\mathcal{N}(0,\mu)$ denotes the one-dimensional normal distribution with mean zero and standard deviation $\mu.$ Throughout the paper, the amplitude of the test signal $Y(t)$ is also reported to maintain perspective of the relative magnitude of the noise.

\vspace{-.2cm}
\section{Numerical Methods}
\label{sec:methods}
\begin{figure}[!p]
    \centering
    \includegraphics[trim={0cm 0.0cm 0cm 0.0cm},clip,scale=0.70]{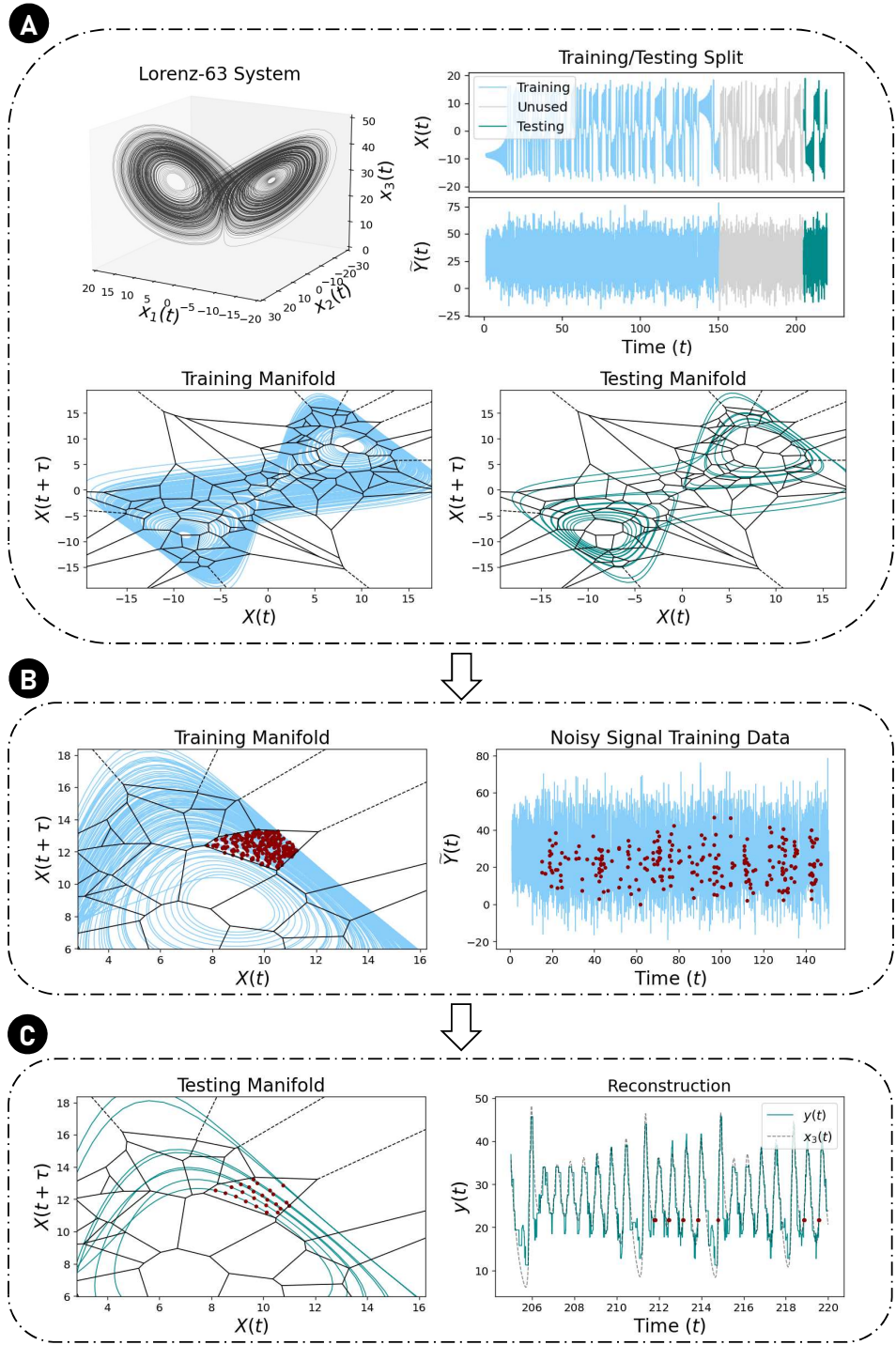}
    \caption{Visualization of the proposed denoising algorithm: In (A), the training and testing data are prepared, in (B) the training procedure is shown by highlighting a set of points on the noisy signal which all correspond with the same training cell, and in (C) the testing process is depicted by indicating the testing points within that cell, as well as their corresponding values on the reconstructed signal. The procedure is explained in detail in \Cref{subsec:datapreparation}, \Cref{subsec:training}, and \Cref{subsec:testing}.}
    \label{fig:flowchart}
\end{figure}

To ameliorate the shortcomings identified in the uniform mesh approach, an unstructured mesh based on a Voronoi diagram is utilized for the signal reconstructions, following the techniques in \cite{araki2021grid}. This results in enhanced signal recovery and faster computation. 
The cell size is adapted according to the data density such that the number of data points per cell is more evenly distributed than in the previously implemented uniform mesh.
\Cref{fig:flowchart} illustrates the three main steps in this nonlinear denoising algorithm: (1) data preparation (\Cref{subsec:datapreparation}), (2) training (\Cref{subsec:training}), and (3) testing (\Cref{subsec:testing}).
Furthermore, \Cref{subsec:refinements} includes improvements beyond the method presented in this section,  (a) using the k-means clustering to refine the Voronoi diagram during the data preparation phase and (b) performing linear interpolation during the testing phase.
From here onwards, the clean reference signal will be referred to as $X(t)$, the noisy/corrupted signal as $\Tilde{Y}(t)$, the true signal of $\Tilde{Y}(t)$ as $Y(t)$, and the denoised/reconstructed signal as $y(t)$. All are sampled at the same frequency from the same dynamical system over the same time interval. Let $N$ be the number of data points in all of the signals and $\Delta t$ be the time-step between sampling times.
\subsection{Data Preparation}
\label{subsec:datapreparation}
Similar to \cite{eckhardt2019spatiotemporal}, both signals $X(t)$ and $\Tilde{Y}(t)$ are split into training and testing data sets. The $R^{\text{th}}$ and $E^{\text{th}}$ data points are used as cutoffs such that all of the data sampled before $t=t_R$ is training data and data sampled after $t=t_E$ is testing data. The length of the training dataset can have implications on the quality of the results in the testing phase, and care should be taken to ensure that the number of training samples is sufficiently large.
It is also worth noting that the samples used for training need not be uniform in time. 
and that the proposed approach is still effective when the training samples are randomly selected from the signal $X(t).$ 

 The time-delay is denoted by $\tau > 0$ and $\tau'= \tau/\Delta t$, which for simplicity is assumed to be an integer, is the corresponding number of time-steps which are delayed. Let $\mathcal{M}^X_{\text{train}}$ and $\mathcal{M}^X_{\text{test}}$ denote the training and testing manifolds, respectively. 
The two manifolds are constructed from the signal $X(t)$ in a dimension $d\geq 2$ by time-delay embedding:
\begin{align*}
    (X(t_i),X(t_i+\tau),X(t_i+2\tau),\dots,X(t_i+(d-1)\tau))&\in\mathcal{M}_{\text{train}}^X, \quad 0\le i < R-(d-1)\tau', \\
    (X(t_j),X(t_j+\tau),X(t_j+2\tau),\dots,X(t_j+(d-1)\tau))&\in\mathcal{M}_{\text{test}}^X, \quad E\le j< N-(d-1)\tau'.
\end{align*}
Throughout, the selection of time-delay $\tau$ is informed by the method in \cite{fraser1986independent,martin2019impact}, which involves identifying the first local minima of the average mutual information and selecting a nearby point that minimizes the number of manifold crossings. Moreover, the choice of embedding dimension $d$ is informed by Cao's method \cite{cao1997practical}. 

A random subset of $\mathcal{M}^X_{\text{train}}$ is chosen to build a k-d tree and construct a Voronoi diagram $V$, which partitions the data in $\mathcal{M}^X_{\text{train}}$ and $\mathcal{M}^X_{\text{test}}$. The first panel of \Cref{fig:flowchart} illustrates these steps with an embedding dimension of $d=2$. 

\noindent A summary of the algorithm is provided below.

\begin{algorithm}[H]
	\SetAlgoLined
	\KwResult{Constructs $\mathcal{M}^{X}_{\text{train}}$, $\mathcal{M}^X_{\text{test}}$, and $V$}
	Let $X_{\text{train}}$ contain $R$ elements of $X$\\
	Let $X_{\text{test}}$ contain $N-E$ elements of $X$\\
	Let $\mathcal{M}^X_{\text{train}}$ be a $d\times (R-\tau' (d-1))$ array\\
	Let $\mathcal{M}^X_{\text{test}}$ be a $d\times (N-E-\tau' (d-1))$ array\\
	\For{$0\leq k<d$}{Assign $\mathcal{M}^X_{\text{train}}[k]$ to be elements $\tau' k$ through $R-\tau' (d-k-1)$ of $X_{\text{train}}$\\
	\vspace{.05cm}
	Assign $\mathcal{M}^X_{\text{test}}[k]$ to be elements $E+\tau' k$ through $N-\tau'(d-k-1)$ of $X_{\text{test}}$
	}
	Take $S$ be a random subsample of $\mathcal{M}^X_{\text{train}}$\\
	Construct a k-d tree from the elements of $S$ to represent $V$
	\caption{Data Preparation}
\end{algorithm}


\subsection{Training Phase}
\label{subsec:training}
If $X(t)$ and $Y(t)$ are causally related signals sampled from the same dynamical system, then by Takens' theorem a diffeomorphic mapping $\mathcal{M}^X  \to \mathcal{M}^Y$ exists. Therefore, there exist mappings $X\to \mathcal{M}^X \to \mathcal{M}^Y \to Y$ given by
$$X(t) \mapsto \underbrace{(X(t),X(t+\tau),\dots,X(t+(d-1)\tau)}_{\in \mathcal{M}^X}\mapsto \underbrace{(Y(t),Y(t+\tau),\dots,Y(t+(d-1)\tau)}_{\in\mathcal{M}^Y}\mapsto Y(t)\vspace{-.2cm}
.$$
However, when the signal $Y(t)$ is corrupted by noise and only $\tilde{Y}(t)$ is available, the manifold $\mathcal{M}^Y$ cannot be accurately constructed and the direct mapping $\mathcal{M}^X \to Y$ is instead sought. For a collection of noisy observations $\{\tilde{Y}(t_i)\}_s$ whose corresponding points $\{(X(t_i),X(t_i+\tau),\dots, X(t+(d-1)\tau))\}$ on $\mathcal{M}^X$ all reside in the same cell $s$ of the Voronoi diagram $V$, the true noise-free observations $\{Y(t_i)\}$ are expected to take on similar values. Thus, by averaging over the observations $\{\tilde{Y}(t_i)\}_s$ in each cell $s$, unwanted noise is reduced and the map $\mathcal{M}^X \to Y$ is efficiently approximated. In the second panel of \Cref{fig:flowchart}, the noisy observations which correspond to the training points of a particular Voronoi cell are highlighted in red. A summary of the algorithm is provided below.

\begin{algorithm}[H]
	\SetAlgoLined
	\KwResult{Assigns cell averages to $V$}
	\For{$0\leq i<R$}{
		Determine the cell $s$ of $V$ the point $\mathcal{M}^X_{\text{train}}(i)$ is in\\
		Append $\tilde{Y}(t_i)$ to a list $A_s$
	}
	Compute the average of each list $A_s$
	\caption{Training Phase}
\end{algorithm}

\subsection{Testing (Reconstruction) Phase}
\label{subsec:testing}
The denoised signal $y(t)$ is reconstructed as follows.
Let $\{t_j\}_{j=E}^{N-(d-1)\tau '-1}$ index the sampling times for the testing data. 
For each $t_j$, a point in phase space is identified according to
\begin{equation*}
    (X(t_j),X(t_j+\tau),X(t_j+2\tau),\dots,X(t_j+(d-1)\tau))\in \mathcal{M}_{\text{test}}^X.
\end{equation*}
The Voronoi cell $s$ that this point resides in is then determined.
Letting $A_s$ denote the average of Voronoi cell $s$, the reconstruction at $t_j$ is obtained by setting $y(t_j)=A_s$. In this case, all of the testing points which belong to the same cell of $\mathcal{M}^X_{\text{test}}$ must be associated with the same value in the signal reconstruction, as illustrated in the third panel of \Cref{fig:flowchart}. Repeating this procedure for all Voronoi cells produces a reconstructed signal, in which the signal can take on exactly as many values as there are Voronoi cells being used. A summary of the testing algorithm is provided below.

\begin{algorithm}[H]
	\SetAlgoLined
	\KwResult{Reconstruction $y(t)$}
	\For{$E\leq j< N$}{
		Determine the cell $s$ of $V$ the point $\mathcal{M}^X_{\text{test}}(j)$ is in\\
		Assign $y(t_j) = A_s$
	}
	\caption{Testing Phase}
\end{algorithm}

\subsection{Algorithm Refinement}
\label{subsec:refinements}
\subsubsection{Interpolation}
\label{subsubsec:interpolation}

\begin{figure}[h]
	\centering
	\includegraphics[trim={0cm 8.0cm 1cm 7.0cm},clip,width=\textwidth]{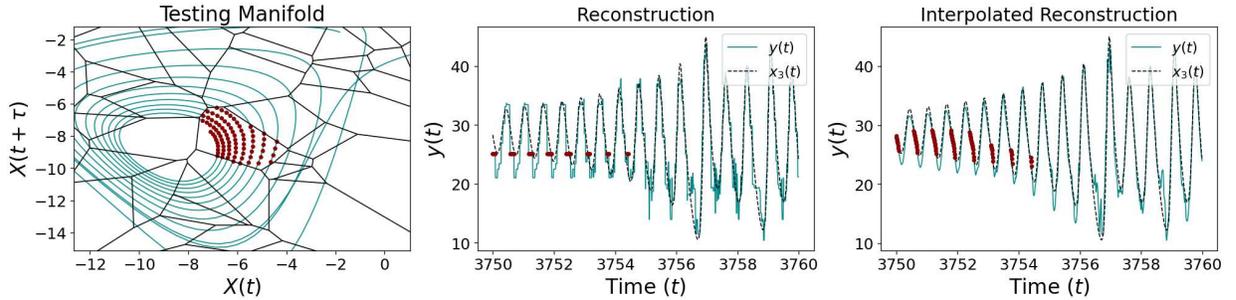}
	\caption{The impact of linear interpolation on the reconstructed signal. The left plot shows a close-up of the Lorenz system's testing manifold and highlights a set of points belonging to a single testing cell in red. The middle plot shows the corresponding value in the reconstructed signal that these points attain without interpolation, and the right plot shows the values with interpolation. Interpolating between values within a cell yields smoother reconstructions.}
	\label{fig:int}
\end{figure}

Reconstructions from the approach described above lead to piecewise constant signal reconstructions which are non-smooth. To address this, interpolation between the Voronoi cell averages is employed, so that the reconstruction $y(t)$ can take on a continuum of values, as opposed to discrete averages $\{A_s\}$ of the Voronoi cells. This improves the regularity of the reconstructed signal from a piecewise constant function to a continuous piecewise linear function. The improvement can be seen in \Cref{fig:int}. 
To implement the interpolation between averages, simplicial elements from the dual graph of the Voronoi diagram, the Delaunay triangulation, are utilized, where each vertex of a simplex is assigned the average of the Voronoi cell it resides in. 
Interpolation then assigns to the point $p$ the average of the weights given to the vertices of the simplex it lies in.
For points that lie outside of the triangulation, the average value associated to the Voronoi cell is used.
\vspace{-.7cm}
\subsubsection{Cell Adaptation: k-Means Clustering}
\label{app:celladaptation}
To improve the algorithm further, $k$-means clustering can be used to construct the unstructured mesh. The $k$-means algorithm iteratively recovers a Voronoi diagram where the variance of the samples within each cell is minimized. By itself, $k$-means clustering is not significantly impactful, but with the addition of linear interpolation, it makes a significant contribution to error reduction as shown in Figure \ref{fig:intercol}.

\section{Results: Application to Lorenz System}
\label{sec:resultsLorenz}
\begin{figure}[!t]
	\centering
	\includegraphics[trim = {.1cm 1.5cm .1cm 1.4cm},clip,width=\textwidth]{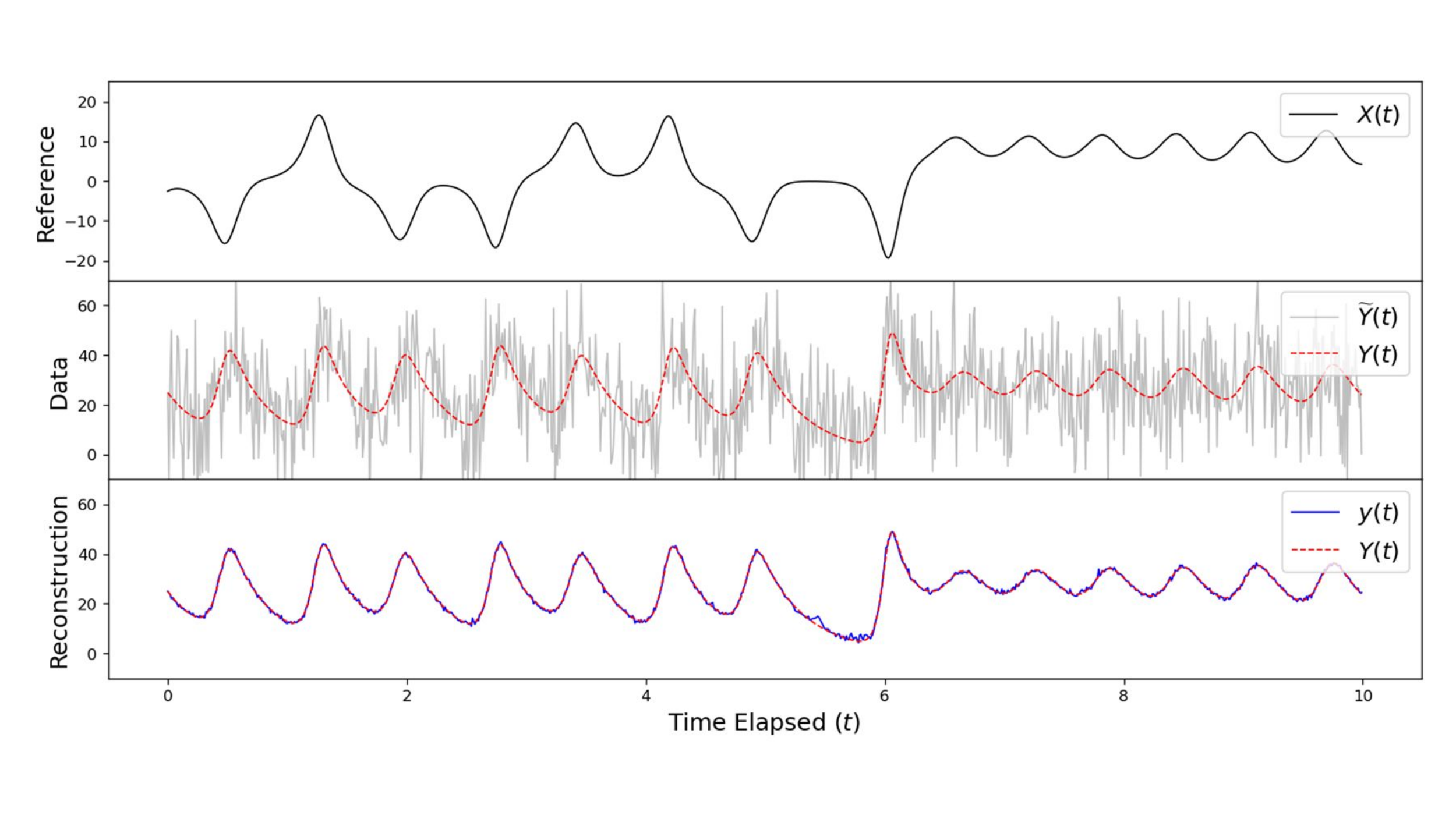}
	\caption{Reconstructing $x_3$ of the Lorenz system from $x_1$, using a time-delay of $\tau=0.17$. The signal $x_3$ is corrupted with Gaussian noise with a standard deviation of $\sigma = 15$, and the target signal $x_3$ has an amplitude of $Y_{\text{max}}-Y_{\text{min}} \approx 43$. An embedding dimension of $d = 3$ along with $10^3$ Voronoi cells are utilized. For training, $5\cdot 10^6$ data points are used, and a gap of $10^6$ timesteps between training and the beginning of testing is left. The experiment took approximately 10 seconds. }
	\label{fig:lor}
	\centering
	\includegraphics[trim = {1.7cm, .5cm, 1.3cm, 0cm},width=\textwidth]{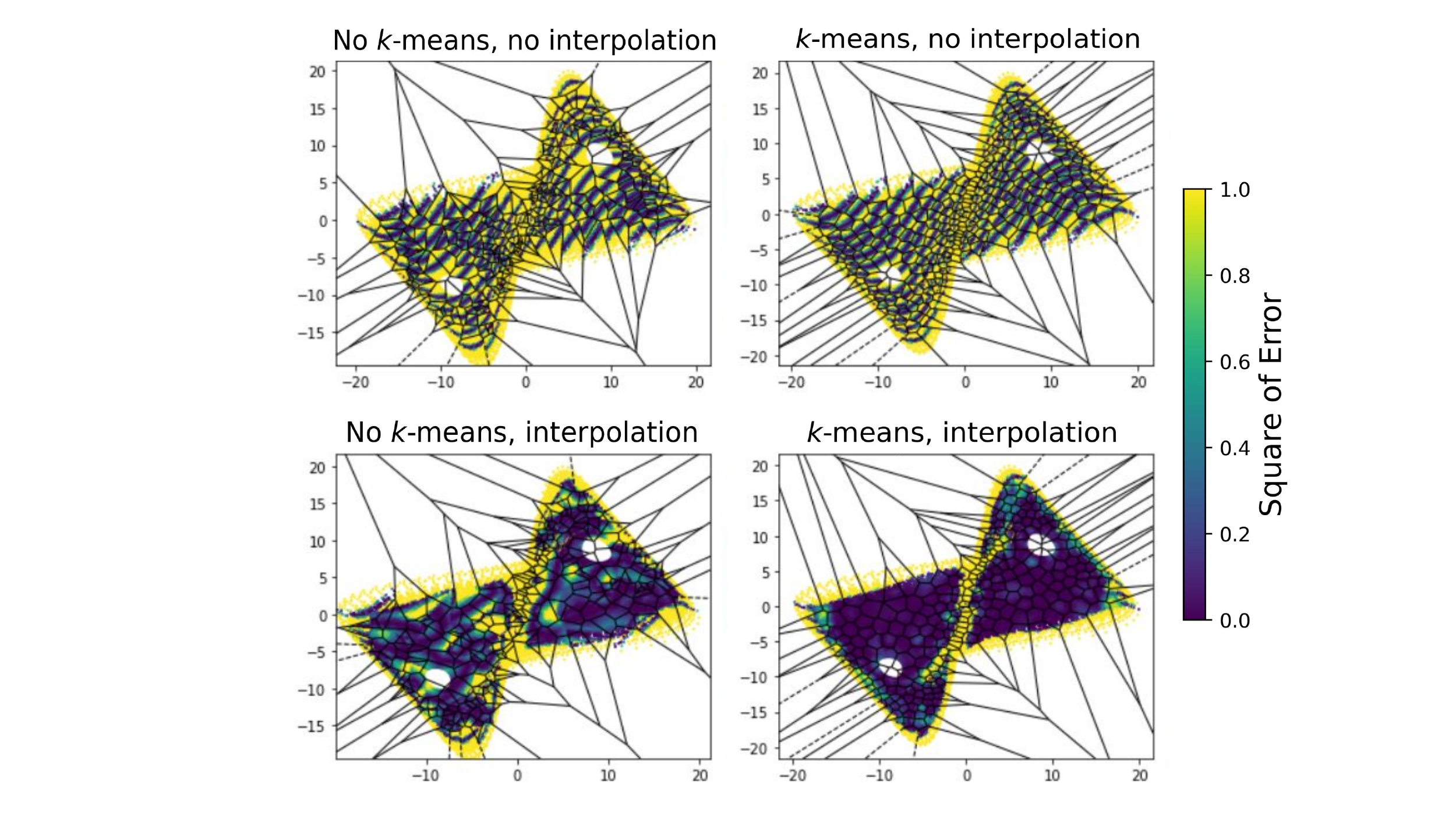}
	\caption{Visualization of the impact of the techniques in \Cref{subsec:refinements}. Shown are four testing manifolds constructed from time-lags of $x_1(t)$ of the Lorenz system. In each case, cell averages are formed using $10^6$ training points to reconstruct $x_3(t)$ and 300 Voronoi cells are used. The error of the resulting reconstruction during testing is then indicated by the brightness of color. }
	\label{fig:intercol}
\end{figure}
The denoising algorithm is first tested on the Lorenz system.
In this study, $x_1(t)$ and $x_3(t)$ of \cref{eqn:Lorenz} are used as the reference signal $X(t)$ and the target signal $Y(t)$, respectively. The corrupted signal $\Tilde{Y}(t)$ is generated by adding Gaussian noise with a standard deviation of $\sigma=15$
to $Y(t)$, and the goal is to denoise $\Tilde{Y}(t)$. In this case, the signal $Y(t)$ has an amplitude of $Y_{\text{max}}-Y_{\text{min}} \approx 43$. \Cref{fig:lor} shows the four signals, where the denoised signal $y(t)$ is obtained using an embedding dimension of $3$.
As shown in \Cref{fig:lor}, the algorithm performed very well, almost completely removing the added noise in the shown testing data.
Minor noise remains in $y(t)$, but it is expected to be reduced as the amount of training data increases and the mesh is refined.

In \Cref{fig:intercol}, the testing manifolds for different reconstructions based upon a two-dimensional embedding are colored to convey the regions of high error. Specifically, the color yellow indicates an area of higher error, while purple indicates an area of lower error. 
The four plots in \Cref{fig:intercol} are obtained with and without the algorithm refinement techniques (linear interpolation and $k$-means clustering) that are covered in \Cref{subsec:refinements}.
As shown in the upper left plot, error is distributed throughout the entire manifold when none of the algorithm refinement techniques are used. 
In \cite{araki2021grid}, this was mitigated by using a smoothing technique in phase space, effectively mixing values of neighboring cells.
Instead, linear interpolation is used here to achieve the same goal, as shown in the bottom left plot. 

Moving on to the bottom right plot of \Cref{fig:intercol}, it is shown that applying the additional technique of $k$-means clustering further improves the reconstruction error. 
Nevertheless, large error still remains near the manifold boundary and on the crossing point.
The large error near the manifold boundary is attributed to the sparse data in the region and is likely improved by using a larger amount of training data.
However, the convergence rate in the region is expected to remain slow due to the low data density.
This might be improved by extrapolating values rather than assigning the cell average values, but this is subject to future work. The system dimension is one higher than the embedding dimension used for the reconstructions in Figure \ref{fig:intercol}, which corresponds to the three dimensional manifold being projected down to two dimensions.
This results in trajectory crossings that are apparent in the shadow manifolds near the origin of Figure~\ref{fig:intercol}.
Therefore, the error in this region is expected to diminish when an embedding dimension of $3$ or higher is used. 

For all remaining signal reconstructions in the paper, linear interpolation without $k$-means clustering is used to permit fast computations. If additional accuracy is required, then the $k$-means algorithm can be used to better distribute the data over the mesh cells. 



\begin{figure}[h!]
    \centering
    \includegraphics[trim={.6cm 4.5cm 0cm 2cm},clip,width=.95\textwidth]{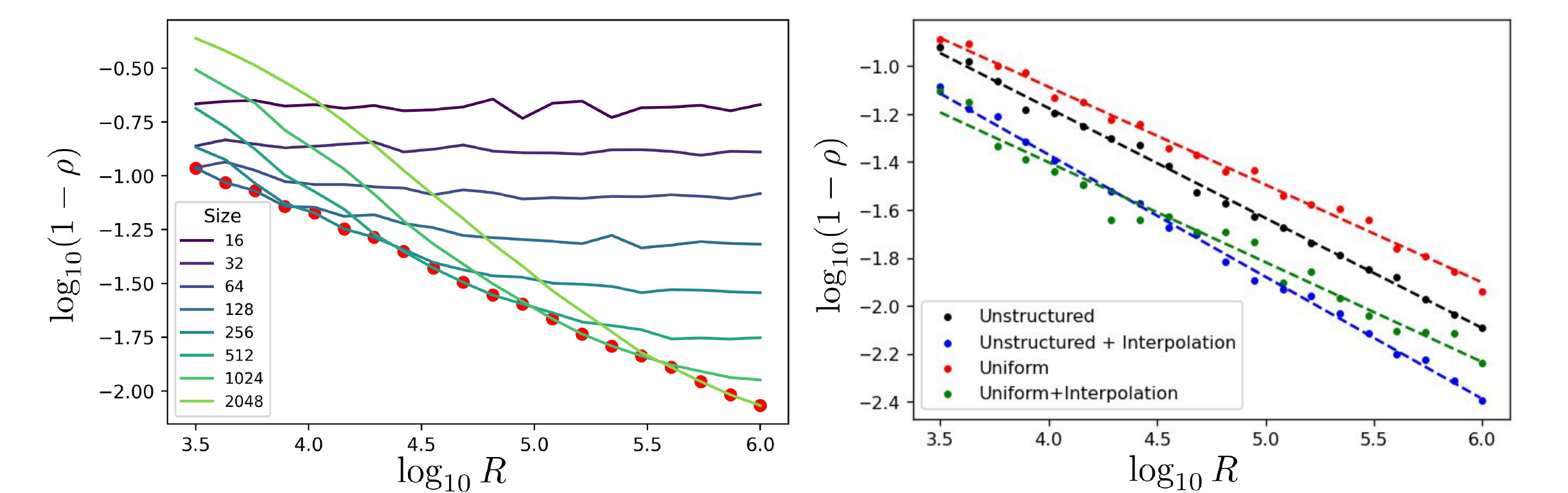}
 \caption{(Left) Error as a function of the number of training points for reconstructing the Lorenz $x_3$ system from $x_1$ for the unstructured mesh without interpolation.  Line colors represent the number of mesh cells. Each point is generated by taking an average of $10$ simulation trials for a fixed amount of training data and mesh size. (Right) Convergence comparison between the uniform and unstructured meshes using optimal numbers of cells from \Cref{table:convergence}. Gaussian noise with a standard deviation of $\sigma = 15$ is used, and the amplitude of the desired $x_3$ signal is roughly 43. Throughout this experiment, the testing phase consists of $10^6$ points.}
    \label{fig:convergence}
\end{figure}

Convergence properties of the method are next obtained for four different cases. Specifically, both the uniform and unstructured meshes are studied with and without linear interpolation for the Lorenz system using an embedding dimension of $d = 3$.
This is accomplished by examining the reconstruction error, $1-\rho$, as a function of the number of training data points, where $\rho$ is the Pearson Correlation Coefficient calculated by \texttt{scipy.stats.pearsonr} \cite{2020SciPy-NMeth}.
The left panel of \Cref{fig:convergence} shows convergence with the amount of training data for different numbers of mesh cells.
From examining linear log-log plots of the error versus the amount of training data, it appears that the error can be made arbitrarily small with a sufficient amount of training data, up to errors induced by aleatory noise and floating point arithmetic. 

The error of the reconstructed signal is largely determined by the number of Voronoi cells.  In other words, for a given dataset, the optimal number of cells depends on the amount of training data and on how noisy the target signal is. For a given amount of training data, the mesh size which produces the lowest-error reconstruction is identified and highlighted in red in \Cref{fig:convergence} (Left). Following one of the curves in \Cref{fig:convergence} (Left), it can also be seen that after an optimal reconstruction error is attained for a fixed mesh size that the convergence rate asymptotes to zero. At this point, the maximum amount of information about the system which can be stored using that number of cells has been reached.

The optimal relationship between the number of mesh cells $N_c$ and amount of available training data $R$ is extracted by performing a linear regression in the log-log plot and fitting to the equation
\begin{equation}
    \log_2 N_c = A\log_{10}R+B,
\end{equation}
where $A$ and $B$ are the fitting parameters, which are given in \Cref{table:convergence} for the four cases.
For the uniform mesh, the number of cells along one dimension is used for $N_c$, whereas for the unstructured mesh $N_c$ denotes the total number of cells. 
The convergence of the denoising algorithm using the optimal relationship between the amount of training data and mesh size is shown in \Cref{fig:convergence} (Right).
It is evident that the slope for the unstructured mesh is larger, even when interpolation is not used.
This is explained by the unstructured mesh's ability to adapt to the density of the available data. Specifically, in regions of low data density, the unstructured mesh approach ensures more samples per cell than the uniform mesh. This is especially important near the edges of the attractor.
Finally, it is shown that linear interpolation greatly improves the error for both meshes.

\begin{table}[h]
  \centering
	\caption{Fitting parameters for the convergence study in \Cref{fig:convergence}}\vspace{-2ex}
	\label{table:convergence}
	\begin{threeparttable}[b]
	\begin{tabularx}{\linewidth}{p{3cm}YYY} \hline\hline
		Mesh & Interpolation & A & B
		\\ \hline 
		Unstructured & No & 1.8 & 0.5 \\
		Unstructured & Yes & 1.7 & 0.9 \\
		Uniform & No & 1.0 & -0.6 \\
		Uniform & Yes & 0.8 & 0.5 \\ \hline \hline
	\end{tabularx}
  \end{threeparttable}
\end{table}
\vspace{-.5cm}



\section{Results: Application to Hall-Effect Thruster System}
\label{sec:resultsHET}
\begin{figure}[h!]
\centering
	\includegraphics[trim={2cm 5.5cm 2.7cm 5.6cm},clip,width=.95\textwidth]{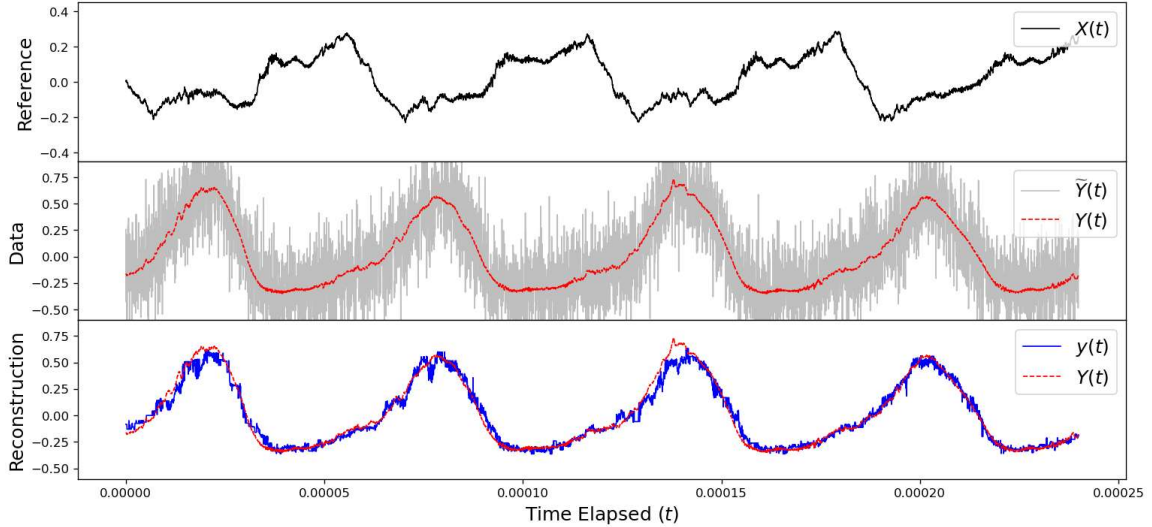}
	\caption{Reconstructing the Cathode Pearson signal from the reference Anode+Cathode. Gaussian noise with a standard deviation of $\sigma = 0.25$ A is added to the Cathode Pearson signal, which has an amplitude of $Y_{\text{max}}- Y_{\text{min}} \approx 1.1$ A. A time-delay of $7\cdot 10^{-6}$ seconds is utilized, along with an embedding dimension of $d = 5$ and $750$ Voronoi cells. For training, $10^5$ data points are used, and a gap of $10^4$ points is left before testing. The experiment took approximately 1 second to run.}
	\label{fig:noisyre}
	\end{figure}

Given the promising convergence results displayed in Figure \ref{fig:convergence}, the denoising algorithm was then applied to several experimentally sampled HET signals.  First, the HET Anode+Cathode signal and the Cathode Pearson signal are used. The Anode+Cathode signal is taken as the reference and a time lag of $7\cdot 10^{-6}$ seconds and embedding dimension of $d = 5$ are chosen.
Gaussian noise with a standard deviation of $\sigma=0.25$ A
is added to the Cathode Pearson signal, which has an amplitude of $Y_{\text{max}}-Y_{\text{min}} \approx 1.1$ A. \Cref{fig:noisyre} shows a reconstruction of the Cathode Pearson signal with the added synthetic noise.
The noise level is significantly reduced in the original signal $\tilde{Y}(t)$, but it appears that some noise still remains in the reconstruction.
	
\begin{figure}[h!]
	\centering
	\includegraphics[trim = {1cm 4cm .5cm 2.2cm},clip,width=.92\textwidth]{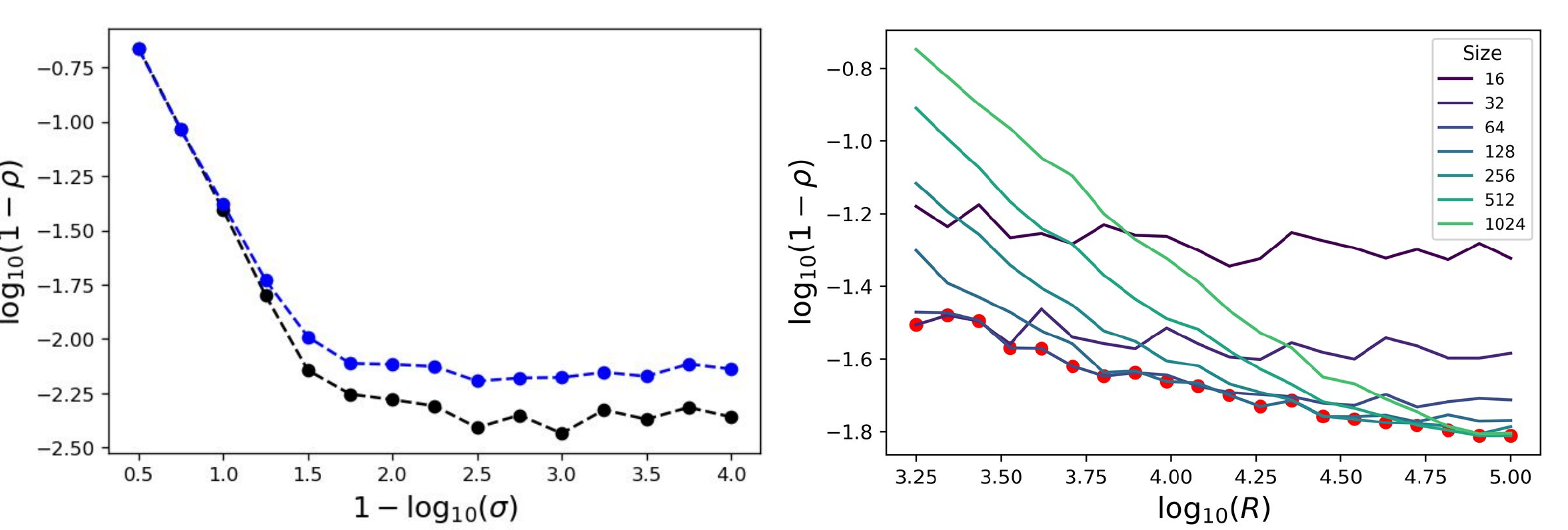}
	\caption{(Left) Reconstruction error with respect to $y'(t)$ (black) and $Y(t)$ (blue) as a function of the synthetic noise level added to the Cathode Pearson data. The abscissa values are $(1-\log_{10} \sigma)$, so the noise level is smaller as the value becomes larger, while the ordinate values are $\log_{10}(1-\rho)$, so the data are better correlated as the value becomes more negative. (Right) Error as a function of the number of training points for reconstrucing the HET Cathode Pearson signal from the Anode+Cathode signal. The chosen values for the time-delay and embedding dimensions are the same as in Figure \ref{fig:noisyre}. Line colors represent the number of mesh cells, and the lowest-error reconstructions for each amount of training data are plotted in red.}
	\label{fig:cmp_err_100k}
\end{figure}

In order to investigate the source of error in the reconstructed signal, the same denoising procedure is applied to the Cathode Pearson data without any additional synthetic noise, $Y(t)$.
Let $y'(t)$ denote the signal reconstructed directly from $Y(t)$, and let $y(t)$ be the denoised signal shown in \Cref{fig:noisyre} reconstructed from $\Tilde{Y}(t)$. The left panel of
\Cref{fig:cmp_err_100k} shows the reconstruction error with respect to $y'(t)$ (black) and $Y(t)$ (blue) as a function of the synthetic noise level added to the Cathode Pearson data.
As the noise level becomes smaller ($\sigma<0.5$ A), the convergence rate starts to taper off, approaching zero. 
This suggests that there is noise in the reference signal not present in the true target signal, limiting the quality of reconstruction and acting as a floor to the reconstruction error.
The gap between the black and blue dots in \Cref{fig:cmp_err_100k} at smaller $\sigma$ indicates the degree of difference between $y'(t)$ and $Y(t)$, which indirectly describes the noise present in the reference signal.
Finally, the steady convergence up to $\sigma=0.5$ A suggests that the algorithm has successfully removed the noise added to the target signal.

\begin{figure}[h!]
    \centering
	\includegraphics[trim={2.5cm 6cm 2.5cm 6.04cm},clip,width=\textwidth]{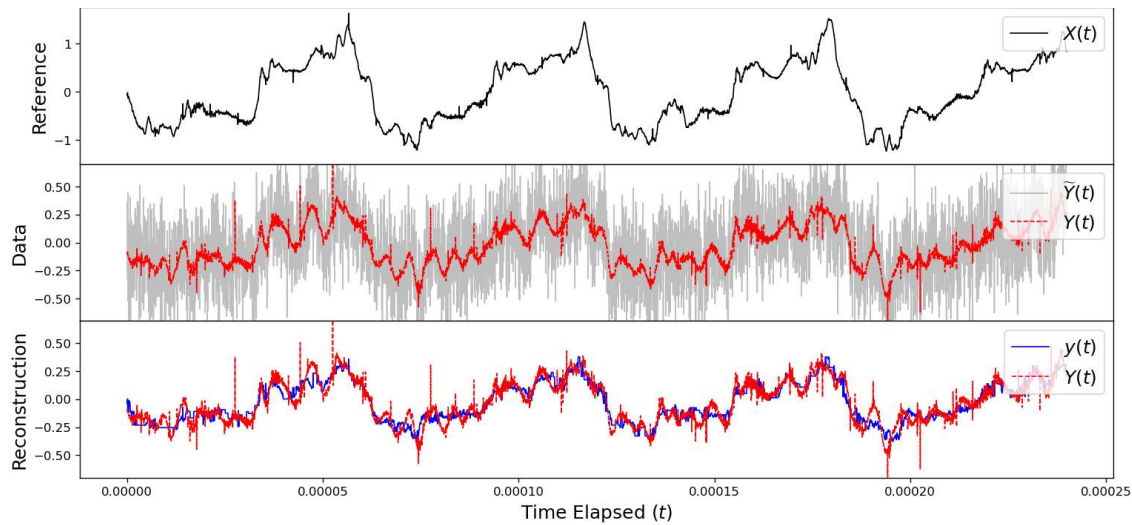}
	\caption{Reconstructing the Ring 6 signal using the Total Cage signal as a reference. Gaussian noise with a standard deviation of $\sigma = 0.25$ A is added to the already noisy Ring 1 signal which has an amplitude of roughly $Y_{\text{max}}-Y_{\text{min}} \approx 1.7$ A. The remaining parameters are the same as in Figure \ref{fig:noisyre}. The experiment took only 1 second of computation time.}
	\label{fig:rings16}
\end{figure}
The right panel of \Cref{fig:cmp_err_100k} shows the error as a function of the number of training points for different numbers of mesh cells. 
The convergence curves are used to determine the optimal sample size (shown in red dots) for each mesh resolution. 
The optimal points are not exactly linear, as in \Cref{fig:convergence}, though this may be attributed in part to the lack of experimental data available far beyond $R=10^5$. 

Finally, the denoising algorithm is applied to data which appears very noisy from the beginning, and where the clean signal without noise is unknown. Moreover, the apparent noise in these measurements cannot be easily distinguished from the real signal representing the dynamics of the system. 
Specifically, the Total Cage current is used as the reference signal, and the current collected at Ring 6 (see Fig.~1 in \cite{eckhardt2019spatiotemporal}) is used as the target signal. 
As shown in \Cref{fig:rings16}, the denoising algorithm appears to work fairly well, removing the highest frequency components not present in the Total Cage current, while retaining some of higher frequency signal.

    

\section{Conclusions}
\label{sec:conclusions}
This paper extended the nonlinear denoising algorithm proposed in \cite{araki2021grid} by utilizing an unstructured mesh which adapts its resolution according to the data density and employing linear interpolation as a means to smooth the reconstructed signal. 
This method assumes availability of a clean reference signal and uses it to reconstruct a noisy signal sampled from the same dynamical system, given that the two signals are strongly causal.
Through the procedure of time-delay embedding, the clean signal is mapped onto phase-space and partitioned by the unstructured mesh cells. 
Then, CCM permits ensemble-averaging of the noisy signal data points which belonging to the same cell grouping. 
Finally, the denoised signal is reconstructed by mapping the phase-space position back to the time-domain and assigning values which are obtained by interpolating between averages.

The denoising method was explored extensively for the Lorenz attractor and a HET system, demonstrating its applicability to quasi-periodic dynamical systems. The efficiency of the proposed approach was also shown, as the signal reconstructions in Figures \ref{fig:lor}, \ref{fig:noisyre}, and \ref{fig:rings16} all required less than ten seconds of computation time. Convergence behaviors were compared for the denoising method with a uniform mesh and an unstructured mesh, and significant improvements were evident when using the unstructured mesh. Such improvements are primarily attributed to the adaptivity of the unstructured mesh, as the number of data points is more evenly distributed across the mesh cells. With lower reconstruction errors and faster statistical convergence than the uniform mesh approach, the unstructured mesh approach to nonlinear noise reduction is a promising tool for studying causally related signals of quasi-periodic dynamical systems.


\section*{Acknowledgments}
\noindent This work was supported in part by AFOSR grant FA-9550-20RQCOR098 (Program Officer: Dr. Frederick Leve). Data used in this paper was obtained from the EPTEMPEST experimental program funded by AFSOR grant FA9550-17QCOR497 (Program Officer: Dr. Brett Pokines). This work was completed during the Research in Industrial Projects for Students program directed by Susana Serna through the UCLA Institute for Pure and Applied Mathematics.
\vspace{-.2cm}
\bibliography{Publication.bbl}

\end{document}